\documentclass[superscriptaddress]{revtex4-1}

\usepackage{subfigure,epsfig,amsmath,graphicx}

\newcommand{\beginfigtable}{%
        \renewcommand{\thetable}{\arabic{table}}%
        \renewcommand{\figurename}{Figure}
        \renewcommand{\thefigure}{\arabic{figure}}%
}

\newcommand{\beginsupplement}{%
        \setcounter{table}{0}
        \renewcommand{\tablename}{Supplementary Table}
        \renewcommand{\thetable}{S\arabic{table}}%
        \setcounter{figure}{0}
        \renewcommand{\figurename}{Supplementary Figure}
        \renewcommand{\thefigure}{S\arabic{figure}}%
}

\begin{document}

\title{Explaining the power-law distribution of human mobility 
\\through transportation modality decomposition}

\author{Kai Zhao}%
\email{kai.zhao@cs.helsinki.fi}
\affiliation{Department of Computer Science, University of Helsinki, Helsinki, Finland}

\author{Mirco Musolesi}%
%\email{m.musolesi@cs.bham.ac.uk}
\affiliation{School of Computer Science, University of Birmingham, Birmingham, UK}

\author{Pan Hui}%
%\email{panhui@cse.ust.hk}
\affiliation{Department of Computer Science and Engineering, 
The Hong Kong University of Science and Technology, Hong Kong}

\author{Weixiong Rao}%
%\email{wxrao@tongji.edu.cn}
\affiliation{School of Software Engineering, Tongji University, Shanghai, China}

\author{Sasu Tarkoma}%
\email{sasu.tarkoma@cs.helsinki.fi}
\affiliation{Department of Computer Science, University of Helsinki, Helsinki, Finland}

\begin{abstract}

Human mobility has been empirically observed to
exhibit L\'{e}vy flight characteristics and behaviour 
with power-law distributed jump size. The fundamental mechanisms behind this behaviour has
not yet been fully
explained. In this paper, we propose to explain
the L\'{e}vy walk behaviour observed in human mobility patterns by
decomposing them into different classes according to the different transportation modes, such as Walk/Run,
Bike, Train/Subway
or Car/Taxi/Bus. 
Our analysis is based on two real-life GPS datasets
containing approximately 10 and 20 million GPS samples with transportation mode information. 
We show that human mobility
can be modelled as a mixture of different transportation modes, and that
these single movement patterns can be approximated by a lognormal distribution
rather
than a power-law distribution.
Then, we demonstrate that the mixture of the decomposed
lognormal flight distributions associated with each modality
is a power-law distribution, providing an explanation to the emergence of L\'{e}vy Walk  
patterns that characterize human mobility patterns. 

\end{abstract}

\maketitle

Understanding human mobility is crucial for epidemic control \cite{Speadone,Speadtwo,Speadthree,Speadfour}, 
urban planning \cite{zhengyuubicomp2011, zhengyukdd2012}, traffic forecasting systems 
\cite{trafficone, traffictwo} and, more recently, various mobile and 
network applications \cite{samulisensys13, transportationAppOne, 
transportationAppTwo, hui2009empirical, TMC14}. 
Previous research has shown that trajectories in human mobility have 
statistically similar features as L\'{e}vy Walks by studying the traces of bank 
notes \cite{banknotes}, cell phone users' locations \cite{celltower} and GPS \cite{LevyTon,WhyLevyOne, WhyLevyTwo, WhyLevyThree}. 
According to the this model, human movement contains many short flights and some long flights, and these flights follow a power-law distribution.  

\begin{figure*}
\centering

\includegraphics[width=0.33\textheight]{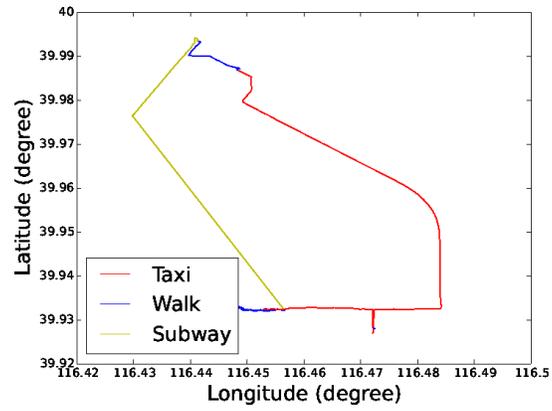}

\caption{Illustration of a synthetic trail (taxi, walk, subway, walk, taxi, walk) for one day trip and their corresponding flights. 
This figure shows that the flight distribution of each transportation mode (walk, taxi, subway) is very different.}
\label{fig:trajectory}
\end{figure*}

Intuitively, these long flights and short flights reflect different transportation modalities. 
Figure. \ref{fig:trajectory} shows a person's one-day trip with 
three transportation modalities in Beijing based on the Geolife dataset (Table \ref{tab:mobilitydataset}) \cite{zhengyugeolife,zhengyuwww09,zhengyuubicomp08}. 
Starting from the bottom right corner of the figure, the person takes a taxi and then walks to the destination
in the top left part.  After two hours, the person takes the subway to another location (bottom left)
and spends five hours there. Then the journey continues and the person takes a taxi back to the
original location (bottom right).  The short flights are associated with walking and the second short-distance taxi trip, 
whereas the long flights are associated with the subway and the initial taxi trip. Based on this simple example, we
observe that the flight distribution of each transportation mode is different.

In this paper, we study human mobility with two large GPS datasets, 
the Geolife and Nokia MDC datasets (approximately 10 million and 20 million GPS samples respectively), 
both containing transportation mode information such as Walk/Run, 
Bike, Train/Subway or Car/Taxi/Bus. 
The four transportation modes (Walk/Run, Bike, Train/Subway and Car/Taxi/Bus) cover the most frequently used human mobility cases. 
First, we simplify the trajectories obtained from the datasets using a rectangular model, 
from which we obtain the flight length \cite{LevyTon}. 
Here a flight is the longest straight-line trip from one point to 
another without change of direction \cite{LevyTon, WhyLevyThree}. One trail 
from an origin to a destination may include several different flights (Fig. \ref{fig:trajectory}). 
Then, we determine the flight length distributions for different 
transportation modes. 
We fit the flight distribution of each transportation mode according to the Akaike information criteria \cite{modelselection} in order to find the best fit distribution.

We show that human movement exhibiting different transportation modalities is better fitted with the lognormal distribution rather than the power-law distribution. 
Finally, we demonstrate that the mixture of these transportation mode distributions is a power-law distribution based on two new findings: first, 
there is a significant positive correlation 
between consecutive flights in the same transportation mode, and second, the elapsed time in 
each transportation mode is exponentially distributed.

The contribution of this paper is twofold.  
First, we extract the distribution function of displacement with different transportation modes. 
This is important for many applications \cite{trafficone, traffictwo, samulisensys13, transportationAppOne, transportationAppTwo}. 
For example, a population-weighted opportunities (PWO) model \cite{PWOmodel} has been developed to predict human mobility patterns in cities. They find that there is a relatively high mobility at the city scale due to highly developed traffic systems inside cities. Our results significantly deepen the understanding of urban human mobilities with different transportation modes. 
Second, we demonstrate that the mixture of 
different transportations can be approximated as a truncated L\'{e}vy Walk. 
This result is a step towards explaining the emergence of L\'{e}vy Walk patterns in human mobility.

\section*{Results} \label{sec:observation}

\begin{figure*}
\centering
\includegraphics[width=0.60\textheight]{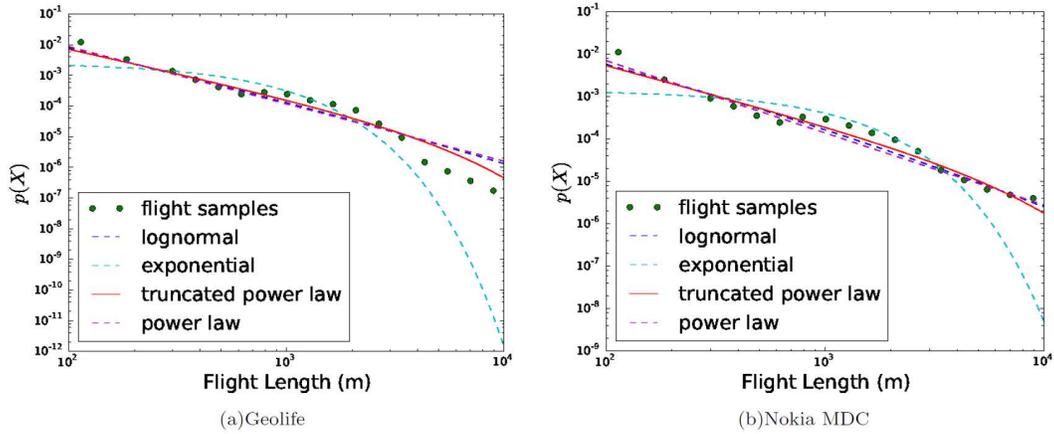}

\caption{Power-law fit for overall flight. 
(a-b) Power-law fitting of all flights regardless of transportation modes in the Geolife and the Nokia MDC dataset . 
The green points refer to the flight length samples obtained from 
the GeoLife and the Nokia MDC dataset, while the solid red line represents 
the best fitted distribution according to Akaike weights. 
The overall flight length distribution regardless of transportation modes 
is well fitted with a truncated power-law distribution.}
\label{fig:all}
\end{figure*}

\textbf{Power-law fit for overall flight.} 
First, we fit the flight length distribution of the  
Geolife and Nokia MDC datasets regardless of transportation modes (see Methods section). 
We fit truncated power-law, lognormal, power-law and exponential distribution (see Supplementary Table S1). 
We find that 
the overall flight length ($l$) distributions fit a truncated power-law $P(l) \propto l^\alpha e ^{\gamma l}$ with 
exponent $\alpha$ as 1.57 in the Geolife dataset ($\gamma = 0.00025$) and 1.39 in the Nokia MDC dataset ($\gamma = 0.00016$) (Fig. \ref{fig:all}), 
better than other alternatives such as power-law, lognormal or exponential. 
Figure. \ref{fig:all} illustrates the PDFs and their best fitted distributions 
according to Akaike weights. The best fitted distribution (truncated power-law) is represented as a solid line and the rest are dotted lines. 
We use logarithm bins to remove tail noises \cite{LevyTon, PowerLawPython}. 
Our result is consistent with previous research (\cite{banknotes,celltower,LevyTon,LevyTon,WhyLevyOne,WhyLevyTwo, WhyLevyThree}), 
and the exponent $\alpha$ is close to their results.

We show the Akaike weights for all fitted distributions in the Supplementary Table S2. 
The Akaike weight is a value between 0 and 1. The larger it is, the better the distribution is fitted \cite{modelselection,PowerLawPython}. 
The Akaike weights of the power-law distributions regardless of transportation modes are 1.0000 in both datasets. 
The p-value is less than 0.01 in all our tests,  which means that our results are very strong in terms of statistical significance. 
Note that here the differences between fitted distributions are not remarkable as shown in the Fig. \ref{fig:all}, especially between the truncated power-law and the lognormal distribution. 
We use the loglikelihood ratio to further compare these two candidate distributions. 
The loglikelihood ratio is positive if the data is more likely in the power-law distribution, and negative if the data is more likely in the lognormal distribution. 
The loglikelihood ratio is 1279.98 and 3279.82 (with the significance value $p < 0.01$) in 
the Geolife and the NokiaMDC datasets respectively, indicating that the data is much better fitted with the truncated power-law distribution.

\begin{figure*}
\centering

\includegraphics[width=0.60\textheight]{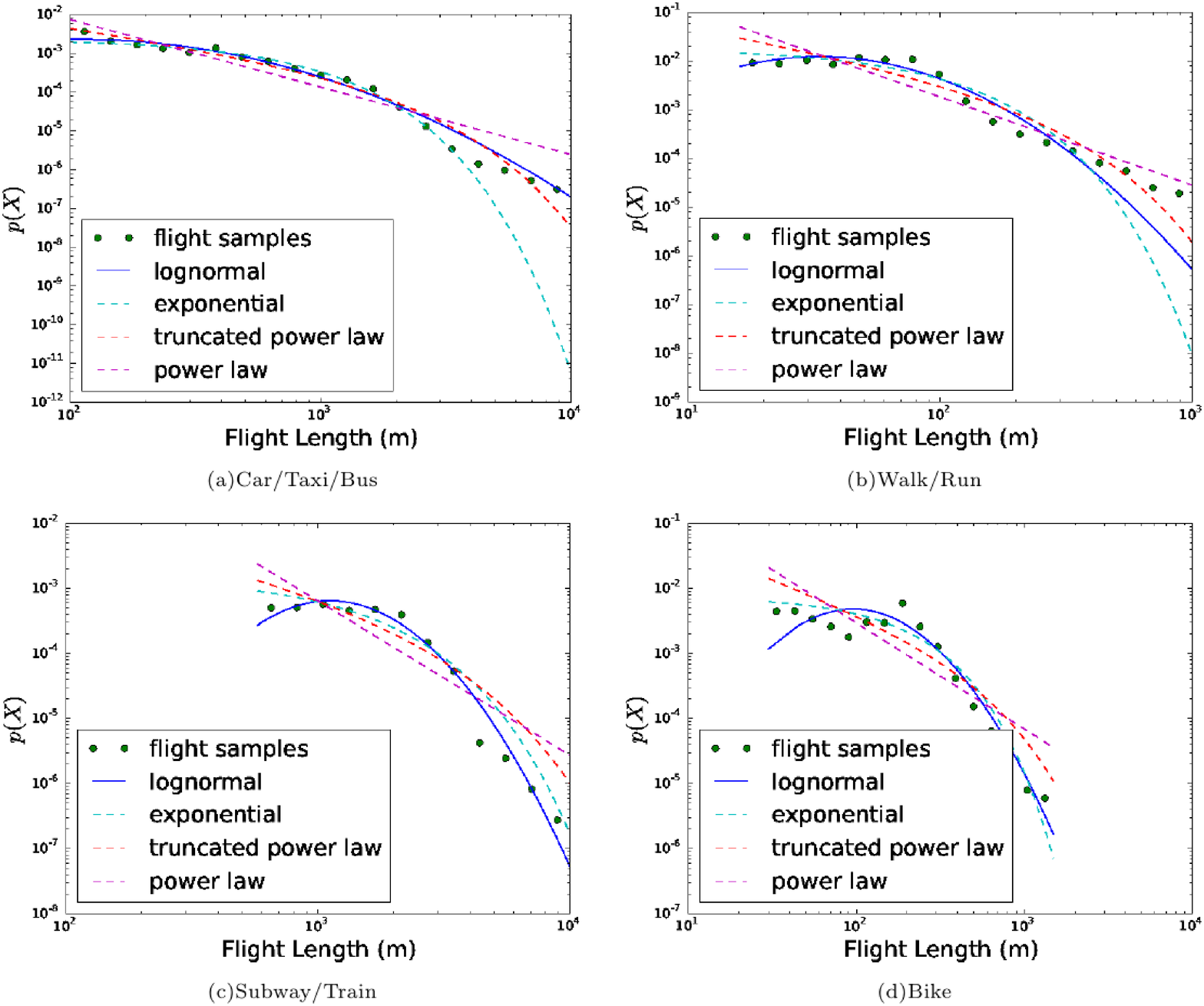}

\caption{Lognormal fit for single transportation mode in the Geolife dataset. 
(a-d) Flight distribution of all transportation modes (Car/Taxi/Bus, 
Walk/Run, Subway/Train, Bike). The green points refer to the flight length samples obtained from 
the GeoLife, while the solid blue line represents 
the best fitted distribution according to Akaike weights. 
The flight length distribution in each transportation mode 
is well fitted with a lognormal distribution.}
\label{fig:geolife}
\end{figure*}

\textbf{Lognormal fit for single transportation mode.} 
However, the distribution of flight lengths in each single transportation mode is not well 
fitted with the power-law distribution. 
Instead, they are better fitted with the lognormal 
distribution (see Supplementary Table S2). 
All the segments of each transportation flight length are best 
approximated by the lognormal distribution with different parameters. 
In Fig. \ref{fig:geolife} and Supplementary Fig. S1, we represent the flight length distributions of Walk/Run, Bike, Subway/Train and Car/Taxi/Bus in 
the Geolife and the Nokia MDC dataset correspondingly. 
The best fitted distribution (lognormal) is represented as a solid line and the rest are dotted lines. 
 
Table \ref{tab:parameters} shows the fitted parameter 
for all the distributions ($\alpha$ in the truncated power-law, $\mu$ and $\sigma$ in the lognormal). 
We can easily find that the $\mu$ is increasing over 
these transportation modes (Walk/Run, Bike, Car/Taxi/Bus and Subway/Train), 
identifying an increasing average distance. 
Compared to Walk/Run, Bike or Car/Taxi/Bus, the flight distribution in Subway/Train mode is more right-skewed, which means that people usually travel to a more distant location by Subway/Train.

It must be noted that our findings for the Car/Taxi/Bus mode are different from these recent research results \cite{ExponentialOne, ExponentialTwo}, which also investigated the case of a single transportation mode, and found that the scaling of human mobility is exponential by examining taxi GPS datasets. 
The differences are mainly because few people tend to travel a long distance by taxi due to economic considerations. 
So the displacements in their results decay faster than those measured in our Car/Taxi/Bus mode cases. 

\textbf{Mechanisms behind the power-law pattern.} 
We characterize the mechanism of the power-law pattern with
L\'{e}vy flights by mixing the lognormal distributions of the transportation modes. Previous research has shown that 
a mixture of lognormal distributions based on an exponential distribution is
a power-law distribution \cite{newman05power,WWWPowerlaw,WWWPowerlaw2,WWWPowerlaw3}. 
Based on their findings, we demonstrate that 
the reason that human movement follows the L\'{e}vy Walk pattern is due to 
the mixture of the transportation modes they take. 

We demonstrate that the mixture of the lognormal distributions of different transportation modes (Walk/Run, Bike, 
Train/Subway or Car/Taxi/Bus) is 
a power-law distribution given two new findings: 
first, we define the change rate as the relative change of length between two consecutive flights with the same transport mode. The change rate in the same transportation mode is small over time. 
Second, the elapsed time between different 
transportation modes is exponentially distributed. 

\textbf{Lognormal in the same transportation mode.} 
Let us consider a generic flight $l_t$. 
The flight length at next interval of time $l_{t+1}$, given the change rate $c_{t+1}$, is 
\begin{equation} 
l_{t+1}=l_t+c_{t+1}l_t.
\end{equation}

\begin{figure*}
\centering

\includegraphics[width=0.60\textheight]{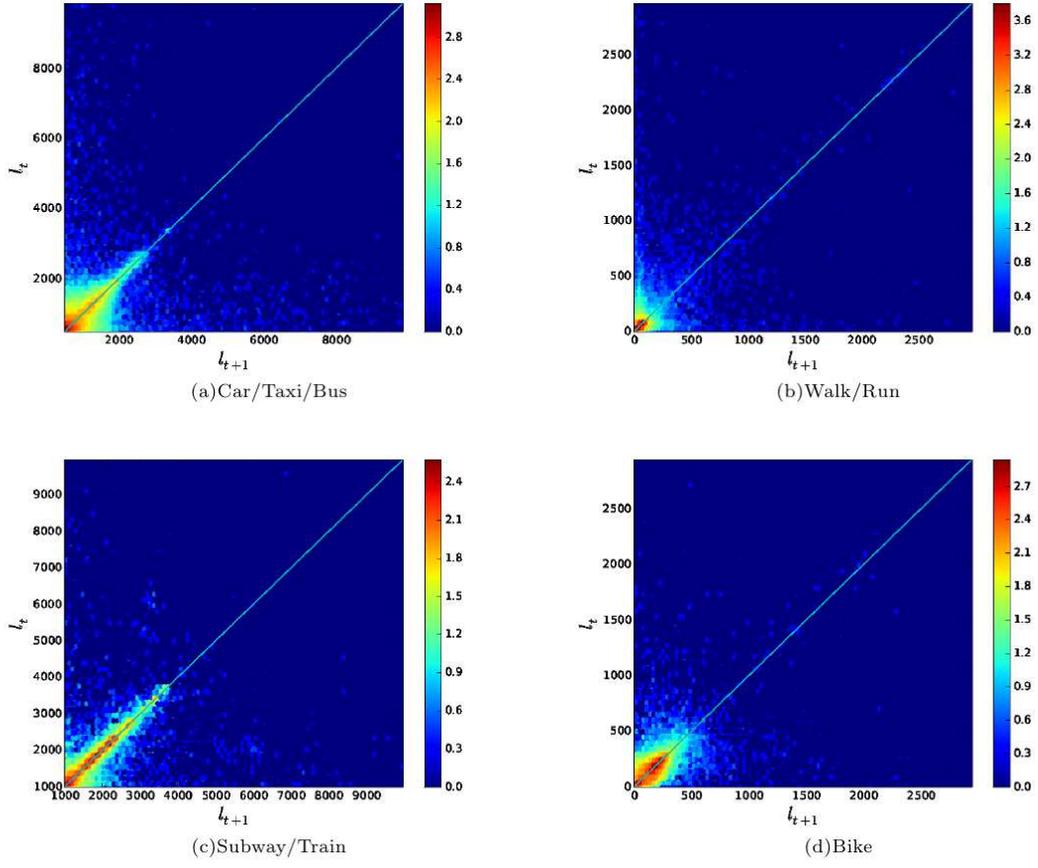}

\caption{Flight length correlation for each transportation mode. 
(a-d) Consecutive Flight length correlation of all transportation modes (Car/Taxi/Bus, 
Walk/Run, Subway/Train, Bike) in the GeoLife dataset. 
A high density of points are near diagonal line $l_t$ = $l_{t+1}$, 
identifying a small difference $l_{t+1}-l_t$ in the same transportation mode between two time steps.}
\label{fig:rate}
\end{figure*}

It has been found that the change rate $c_t$ in the same transportation mode is small over time \cite{samulisensys13, zhengyuubicomp08}. 
The change rate $c_t$ reflects the correlation between two consecutive displacements in one trip. 
To obtain the pattern of correlation between consecutive displacements in each transportation mode, we plot 
the flight length point ($l_t$, $l_{t+1}$) from the GeoLife dataset (Fig. \ref{fig:rate}). Here $l_t$ 
represents the $t$-th flight length and $l_{t+1}$ represents the $t+1$-th flight length in a consecutive trajectory in one transportation mode \cite{Correlation}. 
Figure. \ref{fig:rate} shows 
the density of flight lengths correlation in Car/Taxi/Bus, Walk/Run, Subway/Train and Bike correspondingly. ($l_t$, $l_{t+1}$) are posited near the diagonal line, 
which identifies a clear positive correlation. 
Similar results are also found in the Nokia MDC dataset (see Supplementary Fig. S2).

We use the Pearson correlation coefficient to quantify the strength of the 
correlation between two consecutive flights in one transportation mode \cite{Pearson}. 
The value of Pearson correlation coefficient $r$ is shown in the Supplementary Table S3. 
The $p$ value is less than 0.01 
in all the cases, identifying very strong statistical significances. 
$r$ is positive in each transportation mode and ranges from 0.3640 to 0.6445, 
which means that there is a significant positive correlation 
between consecutive flights in the same transportation mode, and 
the change rate $c_t$ in the same transportation mode between two time steps is small.

The difference $c_t$ in the same transportation mode between two time steps is small due to a small difference $l_{t+1}-l_t$ in consecutive flights. 
We sum all the contributions as follows:

\begin{align} 
\sum_{t=0}^{T}c_t &= \sum_{t=0}^{T} \frac{l_{t+1}-l_t}{l_t} 
\\
& \approx \int_0^T{\frac{dl}{l}} = \ln{\frac{l_T}{l_0}}.
\end{align}

We plot the change rate samples $c_t$ of the Car/Taxi/Bus mode from the Geolife dataset as an example in Supplementary Fig. S3. 
We observe that the change rate $c_t$ fluctuates in an uncorrelated fashion 
from one time interval to the other in one transportation mode 
due to the unpredictable character of the change rate. 
The Pearson correlation coefficient accepts the findings at the 0.03-0.13 level with p-value less than 0.05 (see Supplementary Table S4). 
By the Central Limit Theorem, the sum of the change rate $c_t$ is normally distributed with the mean $\mu T$ and the variance $\sigma^2 T$, 
where $\mu$ and $\sigma^2$ are the mean and variance of the change rate $c_t$ and $T$ is the elapsed time. 
Then we can assert that for every time step $t$, the logarithm of $l$ is also normally distributed with a 
mean $\mu t$ and variance $\sigma^2t$ \cite{formlognormal}.  
Note here that $l_T$ is the length of the flight at the time $T$ after $T$ intervals of elapsed time. In the same transportation mode, the distribution of the flight length with the same change rate mean is lognormal, its density is given by 
\begin{equation} 
P_{singlemode}(l)= \frac{1}{l \sqrt{2\pi \sigma^2t}}exp[-\frac{(\ln(l)-\mu t)^2}{2\sigma^2t}],
\end{equation} 
which corresponds to our findings that in each single transportation mode the flight length is lognormal distributed. 

\textbf{Transportation mode elapsed time.} 
We define elapsed time as the time spent in a particular transportation mode; we found that it is exponentially distributed. 
For example, the trajectory samples shown in Fig. \ref{fig:trajectory} 
contain six trajectories with three different transportation modes, (taxi, walk, subway, walk, taxi, walk). 
Thus the elapsed time also consists of six samples ($t_{taxi1}$, $t_{walk1}$, $t_{subway1}$, $t_{walk2}$, $t_{taxi2}$, $t_{walk3}$). 
The elapsed time $t$ is weighted exponentially between the different transportation modes (see Supplementary Fig. S4). 
Similar results are also reported in \cite{ExponentialOne, ExponentialThree}. 
The exponentially weighted time interval is mainly due to a large portion of Walk/Run flight intervals.
Walk/Run is usually a connecting mode between different transportation modes (e.g., the trajectory samples shown in Fig. \ref{fig:trajectory}), and Walk/Run usually takes much shorter time than any other modes. 
Thus the elapsed time decays exponentially. For example, 87.93$\%$ of the walk distance connecting other transportation modes is within 500 meters and the travelling time is within 5 minutes in the Geolife dataset.

\textbf{Mixture of the transportation modes.} 
Given these lognormal distributions $P_{singlemode}(l)$ in each 
transportation mode and the exponential elapsed time $t$ between different modes, 
we make use of mixtures of distributions. 
We obtain the overall human mobility probability by considering that the 
distribution of flight length is determined by the time $t$, the transportation mode change rate $c_t$ mean $\mu$ and variance $\sigma^2$. 
We obtain the distribution of single transportation mode distribution with the time $t$, the change rate mean $\mu$ and variance $\sigma^2$ fixed. 
We then compute the mixture over the distribution of $t$ since $t$ is 
exponentially distributed over different transportation modes with an exponential parameter $\lambda$. 
If the distribution of $l$, 
$p(l,t)$, depends on the parameter $t$. $t$ is also distributed 
according to its own distribution $r(t)$. Then the 
distribution of $l$, $p(l)$ is given by $p(l) = \int_{t=0}^{\infty} p(l,t)r(t)dt$. Here the $t$ in $p(l,t)$ is the same as the $t$ in the $r(t)$. $r(t)$ is the  exponential distribution of elapsed time $t$ with an exponential parameter $\lambda$.

So the mixture (overall flight length $P_{overall}(l)$) of these lognormal distributions in one 
transportation mode given an exponential elapsed time (with an exponent $\lambda$) between each transportation mode is 
\begin{equation} 
P_{overall}(l)= \int_{t=0}^{\infty} \lambda exp(-\lambda t)\frac{1}{l \sqrt{2\pi \sigma^2t}}exp[-\frac{(\ln(l)-\mu t)^2}{2\sigma^2 t}] d t ,
\end{equation}
which can be calculated to give
\begin{equation} 
P_{overall}(l)= Cl^{-\alpha'} ,
\end{equation}

where the power law exponent $\alpha'$ is determined by $\alpha' = 1- \frac{\mu}{\sigma^2}+\frac{\sqrt{\mu^2+2\lambda\sigma^2}}{\sigma^2}$ 
\cite{WWWPowerlaw,WWWPowerlaw2,WWWPowerlaw3}. 
The calculation to obtain $\alpha'$ is given in Supplementary Note 1. 
If we substitute the parameters presented in Table \ref{tab:parameters}, we will get the $\alpha' = 1.55$ in the Geolife dataset, 
which is close to the original parameter $\alpha = 1.57$, and $\alpha' = 1.40$  in the Nokia MDC dataset, which is close to 
the original parameter $\alpha = 1.39$. 
The result verifies that the mixture of these correlated lognormal distributed flights in one transportation mode given an exponential 
elapsed time between different modes is a truncated power-law distribution.

\section*{Discussion}
 
Previous research suggests that it might be the underlying 
road network that governs the L\'{e}vy flight human mobility, by exploring the human mobility and examining taxi traces in one city in Sweden \cite{WhyLevyThree}. 
To verify their hypothesis, we use a road network dataset of Beijing containing 433,391 roads 
with 171,504 conjunctions and plot the road length distribution \cite{LevyTon, Map1, Map2}. 
As shown in Supplementary Fig. S5, the road length distribution is very different to our 
power-law fit in flights distribution regardless of transportation modes. The $\alpha$ in 
road length distribution is 3.4, much larger than our previous findings $\alpha = 1.57$ in 
the Geolife and $\alpha = 1.39$ in the Nokia MDC. Thus the underlying 
street network cannot fully explain the L\'{e}vy flight in human mobility. 
This is mainly due to the fact that it does not consider many long flights caused by metro or train, and people do not always turn even if they arrive at a conjunction of a
road. Thus the flight length tails in the human mobility should be much larger than those in the road networks.

\section*{Methods}  

\textbf{Data Sets.} 
We use two large real-life GPS trajectory datasets in our work, 
the Geolife dataset \cite{zhengyugeolife} and the Nokia MDC dataset \cite{NokiaMDC}. 
The key information provided by these two datasets is summarized in Table \ref{tab:mobilitydataset}. 
We extract the following information from the dataset: flight lengths and their corresponding transportation modes.
 
Geolife \cite{zhengyugeolife,zhengyuwww09,zhengyuubicomp08} is a public dataset with 
182 users' GPS trajectory over five years (from April 2007 to August 2012) gathered mainly in Beijing, China. 
This dataset contains over 24 million GPS samples with a total distance of 1,292,951 kilometers and a total 
of 50,176 hours. It includes not only daily life routines such as going to work and back home in Beijing, but also 
some leisure and sports activities, such as sightseeing, and walking in other cities. The transportation mode information in this dataset is manually logged by the participants.

The Nokia MDC dataset \cite{NokiaMDC} is a public dataset from Nokia Research Switzerland that aims to study smartphone user behaviour. 
The dataset contains extensive smartphone data of two hundred volunteers in the Lake Geneva region over one and a half years (from September 2009 to April 2011). 
This dataset contains 11 million data points and the corresponding  transportation modes.

\textbf{Obtaining Transportation Mode and The Corresponding Flight Length.} 
We categorize human mobility into four different kinds of transportation modality: 
Walk/Run, Car/Bus/Taxi, Subway/Train and Bike. 
The four transportation modes cover the most frequently used human mobility cases. 
To the best of our knowledge, this article is the first work that 
examines the flight distribution with all kinds of transportation modes in both urban and inter-city environments. 
In the Geolife dataset, users have labelled their trajectories with transportation modes, 
such as driving, taking a bus or a train, riding a bike and walking. 
There is a label file storing the transportation mode labels in each user's folder, 
from which we can obtain the ground truth transportation mode each user is taking 
and the corresponding timestamps. Similar to the Geolife dataset, there is also a file storing the 
transportation mode with an activity ID in the Nokia MDC dataset. 
We treat the transportation mode information in these two datasets as the ground truth.

In order to obtain the flight distribution in each transportation mode, we need to extract the flights. 
We define a flight as the longest straight-line trip from one point to another 
without change of direction \cite{LevyTon,WhyLevyThree}. 
One trail from an original to a destination may include several different 
flights (Fig. \ref{fig:trajectory}). 
In order to mitigate GPS errors, we recompute a position by averaging samples 
(latitude, longitude) every minute. Since people do not necessarily move in perfect straight lines, we need to allow some margin of error in
defining the `straight' line.  We use a rectangular model to simplify the trajectory and obtain the flight length : 
when we draw a straight line between the first point and the last point, the sampled positions between these two endpoints are at a distance less than 10 meters from the line. The same trajectory simplification
mechanism has been used in other articles which investigates the L\'{e}vy walk nature of human mobility \cite{LevyTon}. 
We map the flight length with transportation modes according to 
timestamp in the Geolife dataset and activity ID in the Nokia MDC dataset and 
obtain the final (transportation mode, flight length) patterns. 
We obtain 202,702 and 224,723 flights with transportation mode knowledge in the Geolife and Nokia MDC dataset, respectively.

\textbf{Identifying the Scale Range.} 
To fit a heavy tailed distribution such as a power-law distribution, 
we need to determine what portion of the data to fit ($x_{min}$) and 
the scaling parameter ($\alpha$). 
We use the methods from \cite{newman05power,newman09power} to determine $x_{min}$ 
and $\alpha$. 
We create a power-law 
fit starting from each value in the dataset. Then we select the 
one that results in the minimal Kolmogorov-Smirnov distance, 
between the data and the fit, as the optimal value of $x_{min}$. 
After that, the scaling parameter $\alpha$ in the power-law distribution 
is given by
\begin{equation}
\alpha = 1+ n(\sum^{n}_{i = 1} ln \frac{x_i}{x_{min}})^{-1},
\end{equation}
where $x_i$ are the observed values of $x_i > x_{min}$ and 
$n$ is number of samples.

\textbf{Akaike weights.} 
We use Akaike weights to choose the best fitted distribution. An Akaike weight is a normalized distribution selection criterion \cite{modelselection}.
Its value is between 0 and 1. The larger the value is, the better the distribution is fitted.  

Akaike's information criterion (AIC) is used in combination with Maximum likelihood estimation (MLE). 
MLE finds an estimator of $\hat{\theta}$ that maximizes the likelihood function $L(\hat{\theta}|data)$ 
of one distribution. 
AIC is used to describe the best fitting one among all fitted distributions, 
\begin{equation} 
AIC = -2 log \left(L(\hat{\theta}|data)\right) + 2K.
\end{equation}
Here K is the number of estimable parameters in the approximating model.

After determining the AIC value of each fitted distribution, we normalize these values as follows. 
First of all, we extract the difference between different AIC values called $\Delta_i$, 
\begin{equation}
\Delta_i = AIC_i - AIC_{min}. 
\end{equation}

Then Akaike weights $W_i$ are calculated as follows,
\begin{equation} 
W_i = \frac{exp(-\Delta_i / 2)}{\sum_{r = 1}^{R} exp(-\Delta_i / 2)}.
\end{equation}

\section*{Author contributions} 
K.Z., M.M., P.H., W.R. and S.T. designed the research based on the initial idea by K.Z. and S.T..
K.Z. executed the experiments guided by M.M., P.H., W.R. and S.T. 
K.Z. and S.T. performed statistical analyses, and prepared the figures. 
K.Z., M.M., P.H., W.R. and S.T. wrote the manuscript. 
All authors reviewed the manuscript.

\section*{Additional Information}
\textbf{Competing financial interests:} The authors declare no competing financial interests.

\beginfigtable

\clearpage

\clearpage

\clearpage

\clearpage

\clearpage

\begin{table}
    \centering
    \begin{tabular}{ c c c }
    \hline
    &  Geolife & Nokia MDC \\ \hline
   Location & Beijing & Geneva \\
   Measurement & GPS & GPS \\ 
   Number of samples & 24,876,978 & 11,077,061 \\ 
   Duration & 5 years & 1.5 year \\ 
   Accuracy & 3 $m$  & 3 $m$ \\ 
   Sampling interval& 1-5s & 10s \\ 
   Number of participants & 182 & 200 \\ 
   Number of flights with  & 202,702 & 224,723 \\ 
   transportation mode & & \\
   \end{tabular}
    \caption{The Geolife and the Nokia MDC Human Mobility Datasets.}
\label{tab:mobilitydataset}
\end{table}

\begin{table*}
    \centering
    \begin{tabular}{  c c  c c c c c  c}
    \hline
 Dataset & Transportation Mode & Fitted Distribution &  p & Parameters \\ \hline
     & Overall  &  Truncated Power-law & 0.00 & $\alpha = 1.57$, $\gamma = 0.00025$\\ 
     &Walk/Run  &  Lognormal & 0.00 &$\mu = 4.08$, $\sigma =0.76 $ \\ 
GeoLife    & Bike  &  Lognormal  & 0.00 &$\mu = 5.03$, $\sigma = 0.68$  \\ 
    & Car/Bus/Taxi   &  Lognormal & 0.00&$\mu = 5.78$, $\sigma = 1.04 $\\ 
     & Subway/Train &  Lognormal & 0.00 &$\mu = 7.27$, $\sigma = 0.51$  \\ \hline
     & Overall  &  Truncated Power-law &0.00 & $\alpha = 1.39$, $\gamma = 0.00016$\\ 
    & Walk/Run  &  Lognormal &0.00 & $\mu = 4.58$, $\sigma = 1.09$\\ 
Nokia MDC   & Bike  &  Lognormal &0.00 & $\mu = 5.80$, $\sigma = 1.08$\\ 
   & Car/Bus/Taxi  &  Lognormal &0.00 & $\mu = 6.89$, $\sigma = 0.91$ \\ 
     & Subway/Train  &  Lognormal &0.00 & $\mu = 6.93$, $\sigma = 0.94$\\ 
\hline
 \end{tabular}
    \caption{Parameters of fitted distributions in the GeoLife and in the Nokia MDC datasets. 
    The p-value is less than 0.01 in all the fitted distributions, identifying a strong statistical significance.}
\label{tab:parameters}
\end{table*}

\clearpage

\section*{Supplementary Information for
\\
Explaining the Power-law Distribution of Human Mobility Through Transportation Modality Decomposition} 

\subsection*{Kai Zhao, Mirco Musolesi, Pan Hui, Weixiong Rao, Sasu Tarkoma}

\clearpage

\beginsupplement

\beginsupplement

\begin{figure}
\centering
\includegraphics[width=0.60\textheight]{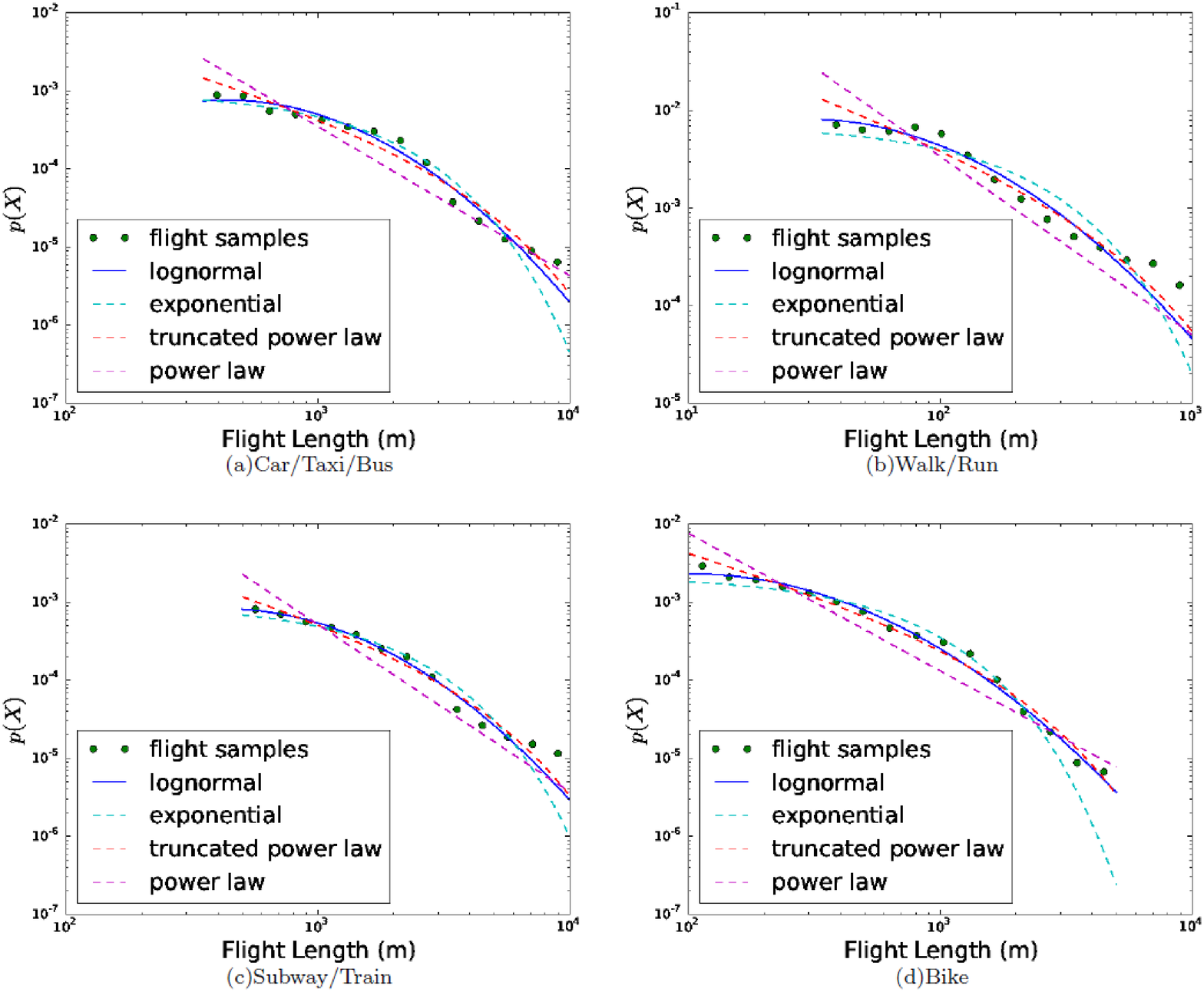}

\caption{Lognormal fit for single transportation mode in the Nokia MDC dataset. 
(a-d) Flight distribution of all transportation modes (Car/Taxi/Bus, 
Walk/Run, Subway/Train, Bike). The green points refer to the flight length samples obtained from 
the Nokia MDC dataset, while the solid blue line represents 
the best fitted distribution according to Akaike weights. 
The flight length distribution in each transportation mode 
is well fitted with a lognormal distribution.}
\end{figure}
\clearpage

\begin{figure*}
\centering

\includegraphics[width=0.60\textheight]{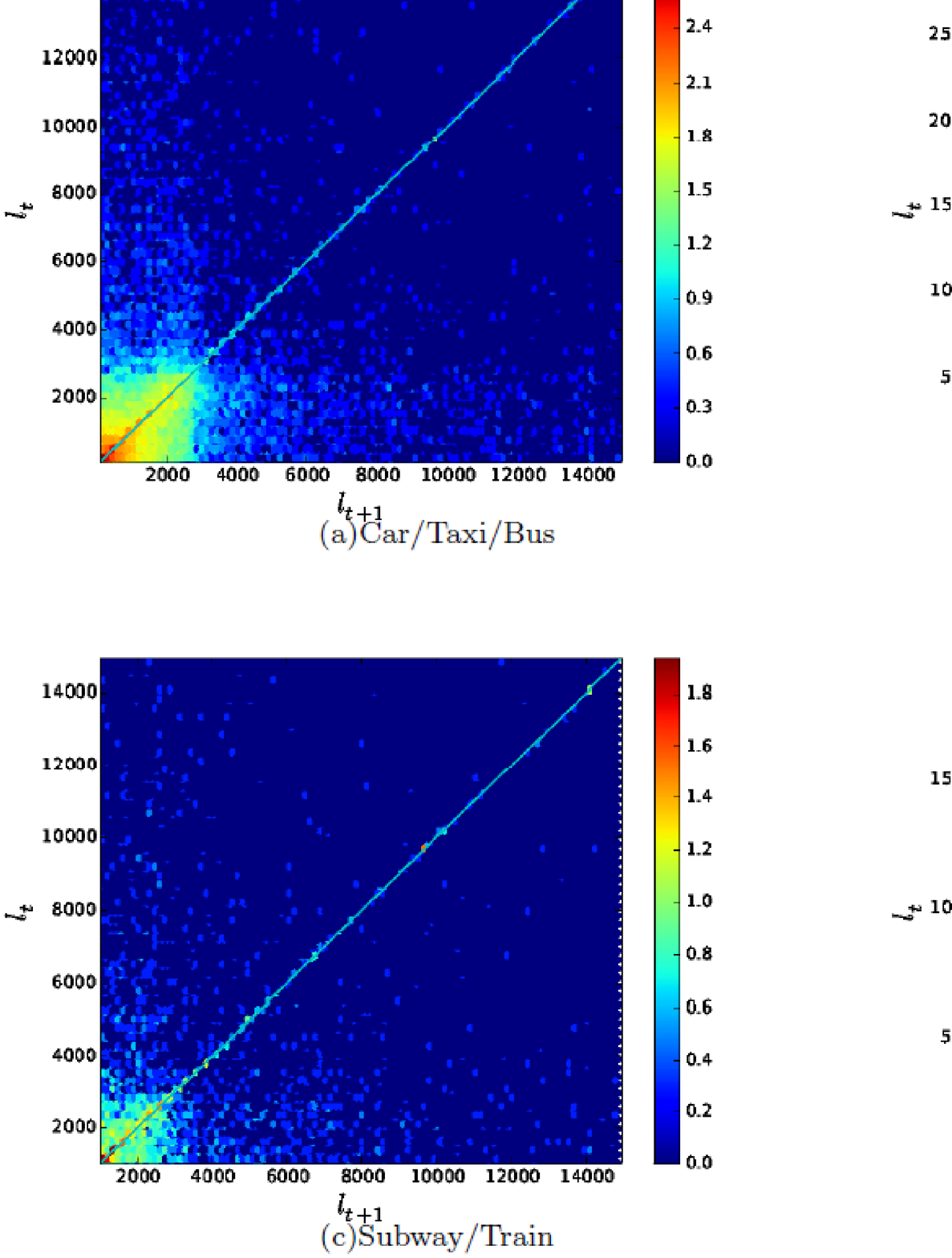}

\caption{Flight length correlation for each transportation mode. 
(a-d) Consecutive Flight length correlation of all transportation modes (Car/Taxi/Bus, 
Walk/Run, Subway/Train, Bike) in the Nokia MDC dataset. 
A high density of points are near diagonal line $l_t$ = $l_{t+1}$, 
identifying a small difference $l_{t+1}-l_t$ in the same transportation mode between two time steps.}

\end{figure*}

\clearpage

\begin{figure*}
\centering

\includegraphics[width=0.33\textheight]{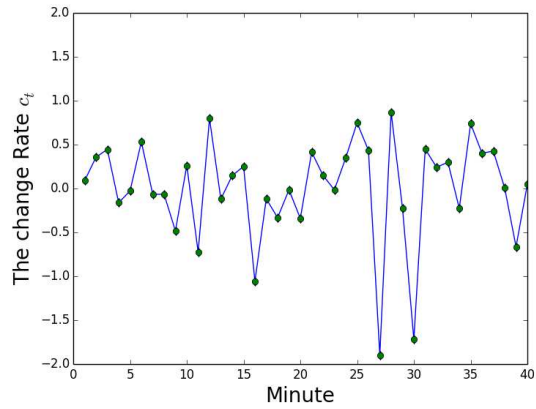}

\caption{The change rate of the Car/Taxi/Bus mode in the Geolife dataset. 
The change rate is defined as the relative change of length between two consecutive flights with the same transport mode. 
From the figure we observe that the change rate are uncorrelated from one time interval to the other.}

\end{figure*}

\begin{figure}
\centering
\includegraphics[width=0.60\textheight]{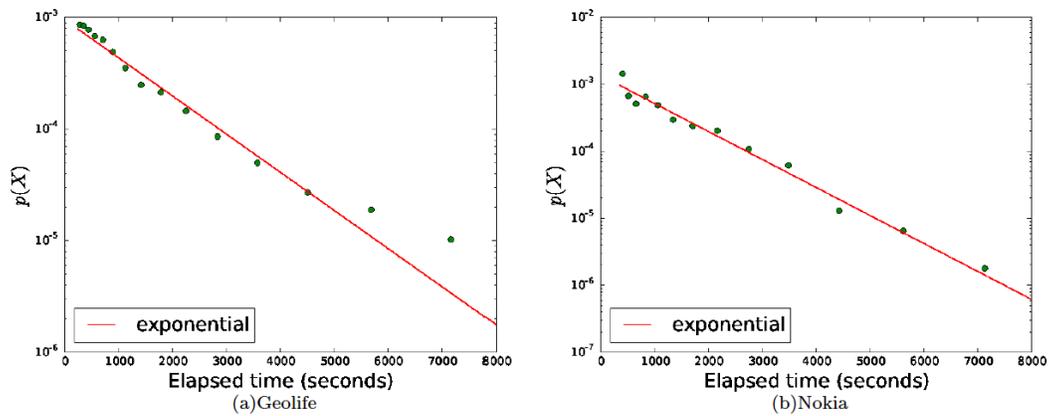}

\caption{Exponential elapsed time. 
The elapsed time $t$ is weighted exponentially between the different transportation modes. 
The exponentially weighted time interval is mainly due to a large portion of Walk/Run flight intervals.}
\end{figure}

\clearpage

\begin{figure}
\centering
\includegraphics[width=0.33\textheight]{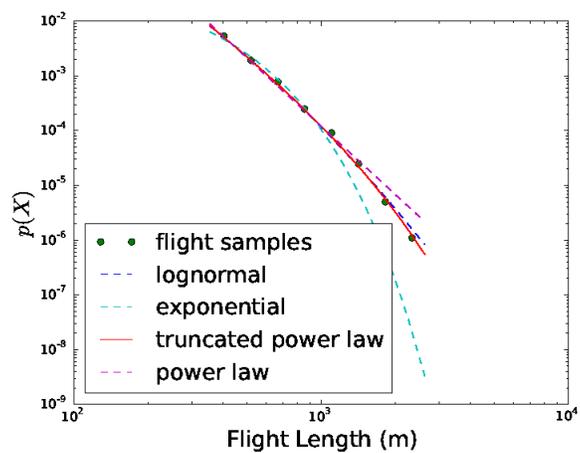}
\caption{The power-law distribution of street length. 
This figure shows that the fitted road length distribution is 
very different to our 
truncated power-law fit in flights distribution. 
The flight length tails in the human mobility is much larger than those shown in this figure.}
\end{figure}

\clearpage

\begin{table}
    \centering
    \begin{tabular}{  c c }
    \hline 
   Distribution & Probability density function (pdf) \\ \hline
   Truncated power-law &  $C x^{-\alpha}e^{-\gamma x}$ \\ 
   Lognormal & $\frac{1}{x\sigma \sqrt{2\pi}}exp[-\frac{(ln(x)-\mu)^2}{2\sigma^2}]$    \\ 
   Power-law & $  C x^{-\alpha}$   \\ 
   Exponential & $\lambda e^{-\lambda x}$    \\ 
   \end{tabular}
    \caption{Fitted distributions.}
\end{table}

\begin{table}
    \centering
    \begin{tabular}{ c c c c c}
    \hline
  Geolife & Truncated & Lognormal & Power-law & Exponential\\ 
   & Power-law & &  & \\ \hline

 Overall  &  1.0000 & 0.0000 & 0.0000 & 0.0000\\ 
 Car/Bus/Taxi  &  0.0000 & 1.0000 & 0.0000 & 0.0000\\ 
 Subway/Train  &  0.0000 & 1.0000 & 0.0000 & 0.0000\\ 
 Walk/Run  &  0.0000 & 1.0000 & 0.0000 & 0.0000\\ 
 Bike  &  0.0000 & 1.0000 & 0.0000 & 0.0000\\ \hline
  Nokia MDC & Truncated & Lognormal & Power-law & Exponential\\ 
   & Power-law & &  & \\ \hline
 Overall  &  1.0000 & 0.0000 & 0.0000 & 0.0000\\ 
 Car/Bus/Taxi  &  0.0000 & 1.0000 & 0.0000 & 0.0000\\ 
 Subway/Train  &  0.0000 & 1.0000 & 0.0000 & 0.0000\\ 
 Walk/Run  &  0.0000 & 1.0000 & 0.0000 & 0.0000\\ 
 Bike  &  0.0000 & 1.0000 & 0.0000 & 0.0000\\
 \end{tabular}
    \caption{Akaike weights of fitted distributions in the Geolife and the Nokia MDC datasets.}
\end{table}

\begin{table}
    \centering
    \begin{tabular}{ c c c }
    \hline 
  Geolife & Pearson correlation $r$  & $p$ \\ \hline
     Car/Bus/Taxi & 0.3640 & 0.0000 \\ 
   Subway/Train & 0.6445 & 0.0000 \\ 
  Walk/Run & 0.5402 & 0.0000 \\ 
   Bike & 0.5584 & 0.0000 \\ 
  Nokia MDC & Pearson correlation $r$ & $p$ \\ \hline
   Car/Bus/Taxi & 0.3980 &  0.0000 \\ 
   Subway/Train & 0.4681 & 0.0000 \\ 
  Walk/Run & 0.4570 & 0.0000 \\ 
   Bike & 0.5291 & 0.0000 \\ 
   \end{tabular}
    \caption{Pearson correlation coefficient for consecutive flights length in the Geolife and the Nokia MDC dataset.}
\end{table}

\clearpage

\begin{table}
    \centering
    \begin{tabular}{ c c c }
    \hline 
  Geolife & Pearson correlation $r$  & $p$ \\ \hline
     Car/Bus/Taxi & 0.0669 & 0.0000 \\ 
   Subway/Train & 0.0705 & 0.0000 \\ 
  Walk/Run & 0.1342 & 0.0000 \\ 
   Bike & 0.1121 & 0.0000 \\ 
  Nokia MDC & Pearson correlation $r$ & $p$ \\ \hline
   Car/Bus/Taxi & 0.0292 &  0.0001 \\ 
   Subway/Train & 0.0282 & 0.0373 \\ 
  Walk/Run & 0.0596 & 0.0000 \\ 
   Bike & 0.1288 & 0.0020 \\ 
   \end{tabular}
    \caption{Pearson correlation coefficient for the change rate in the Geolife and the Nokia MDC dataset.}
\end{table}

\clearpage

\section*{Supplementary Note 1}

Given
\begin{equation} 
P(x)= \int_{t=0}^{\infty} \lambda exp(-\lambda t)\frac{1}{x \sqrt{2\pi \sigma^2t}}exp[-\frac{(\ln(x)-\mu t)^2}{2\sigma^2 t}] d t .
\end{equation}

The calculation to obtain $\alpha'$ is as follows,

\begin{align*} 
P(x)
&= \int_{t=0}^{\infty} \lambda exp(-\lambda t) \frac{1}{x\sigma \sqrt{2\pi t}}exp[-\frac{(\ln(x)-\mu t)^2}{2\sigma^2 t}] d t 
\\
&= \frac{\lambda}{\sigma}\frac{1}{\sqrt{2\pi}} x^{-1}
\\&
\int_{t=0}^{\infty} exp(-\lambda t) exp[-\frac{(\ln(x)-\mu t)^2}{2t\sigma ^2}]
\frac{1}{\sqrt{t}}]dt
\\
&= \frac{\lambda}{\sigma}\frac{1}{\sqrt{2\pi}} x^{-1}
\\&
\int_{t=0}^{\infty} exp[\frac{-(\ln(x)-\mu t)^2 - 2\lambda\sigma^2 t}
{2t\sigma ^2}]
\frac{1}{\sqrt{t}}]d t
\\
&= \frac{\lambda}{\sigma}\frac{1}{\sqrt{2\pi}} x^{-1}
exp(\frac{\ln{x} \mu}{\sigma^2})
\\
&\int_{t=0}^{\infty} exp[-(\frac{\mu^2+2\lambda\sigma^2}{2\sigma^2})t - 
\frac{(\ln x)^2}{2\sigma^2}\frac{1}{t}]
\frac{1}{\sqrt{t}}]d t .
\end{align*}
Using the substitution $t = u^2$ gives
\begin{align*}
P (x)
&= \frac{\lambda}{\sigma}\frac{1}{\sqrt{2\pi}} x^{-1}
exp(\frac{\ln{x} \mu}{\sigma^2})
\\
&
\int_{u=0}^{\infty} exp[-(\frac{\mu^2+2  \lambda \sigma^2}{2\sigma^2})u^2 - 
\frac{(\ln x)^2}{2\sigma^2}\frac{1}{u^2}]
\frac{1}{\sqrt{u^2}}] 2 u du .
\end{align*}
Let $a=\frac{\mu^2+2  \lambda \sigma^2}{2\sigma^2} $ and 
$b=(\ln x)^2{2\sigma^2}$, from the integral table we get
\begin{align*}
\int_{u=0}^{\infty}exp(-a u^2-\frac{b}{u^2}) = 
\frac{1}{2}\sqrt{\frac{\pi}{a}} exp(-2\sqrt{ab}) ,
\end{align*}
which helps us to get the expression for $P(x)$,
\begin{align*}
P(x)
&=
\frac{\lambda}{\sigma \sqrt{\frac{\mu^2}{\sigma^2}-2\lambda^2}}
x^{-(1-\frac{\mu}{\sigma^2}+\frac{\sqrt{\mu^2 +2 \lambda\sigma^2}}{\sigma^2})}
\\
&= \frac{\lambda}{\sigma \sqrt{\frac{\mu^2}{\sigma^2}-2\lambda^2}}x^{-\alpha'}.
\end{align*}
The expression for $\alpha'$ is 
\begin{align*}
\alpha' = 1- \frac{\mu}{\sigma^2}+\frac{\sqrt{\mu^2+2\lambda\sigma^2}}{\sigma^2}.
\end{align*}

Here the $\mu$ and the $\sigma^2$ are the normalized mean and variance of the change rate, while the $\lambda$ is the exponential parameter of 
elapsed time between different transportation modes. We normalize the $\mu$ and $\sigma^2$ of different transportation modes following $\mu = (\mu_{Car/Bus/Taxi} + \mu_{Subway/Train} + \mu_{Walk/Run} + \mu_{Bike}) / 4$ and 
$\sigma^2 = (\sigma_{Car/Bus/Taxi}^2 + \sigma_{Subway/Train}^2 + \sigma_{Walk/Run}^2 + \sigma_{Bike}^2) / 4$. 
Note here $\mu_{Car/Bus/Taxi}$, $\mu_{Subway/Train}$, $\mu_{Walk/Run}$, $\mu_{Bike}$ and 
$\sigma_{Car/Bus/Taxi}$, $\sigma_{Subway/Train}$, 
$\sigma_{Walk/Run}$, $\sigma_{Bike}$ represent the mean and standard deviation of the change rate in each transportation modes in both datasets, as shown in the Table. 2. 
The mean value $\mu$ is 5.54 and 6.05 and 
the variance $\sigma^2$ is 0.5954 and 1.0165 in the Geolife dataset and in the Nokia MDC dataset respectively. Combining the fitted exponential parameter $\lambda =3.16$ in the Geolife dataset and $\lambda = 2.53$ in the Nokia MDC dataset, 
we obtain the final $\alpha' = 1.55$ in the Geolife dataset, 
which is close to the original parameter $\alpha = 1.57$, and $\alpha' = 1.40$  in the Nokia MDC dataset, which is close to 
the original parameter $\alpha = 1.39$. 

\end{document}